\documentclass[prd,preprint,nofootinbib]{revtex4}

\preprint{EPHOU-24-014} 
\usepackage{hyperref}
\usepackage{graphics} 
\usepackage[dvipdfmx]{graphicx}
\usepackage{amsfonts}
\usepackage{latexsym}
\usepackage{amsmath}
\usepackage{amssymb}

\begin{document}


\title{Discovery potential of a long-lived partner of inelastic dark matter at MATHUSLA in $U(1)_{X_3}$ extension of the standard model  }
 
\author{Nobuchika Okada}
\address{Department of Physics and Astronomy, University of Alabama, Tuscaloosa, Alabama 35487, USA}\email{okadan@ua.edu} 
\author{Osamu Seto}
\address{Department of Physics, Hokkaido University, Sapporo 060-0810, Japan} 
 \email{seto@particle.sci.hokudai.ac.jp} 

%

\begin{abstract}
We investigate the discovery potential at the MATHUSLA experiment
 of a long-lived particle (LLP), which is the heavier 
 state of inelastic scalar dark matter (DM)
 in third-generation-philic $U(1)$ ($U(1)_{X_3}$) extension of the Standard Model.
Since the heavier state and DM state form the complex scalar charged under the $U(1)_{X_3}$,
 it is natural that the heavier state $P$ is almost degenerate with the DM state
 and is hence long-lived.
We find that third generation-philic right-handed $U(1)$, $U(1)_{R_3}$, model is the most interesting, because third-generation-philic models are less constrained by the current experimental results and right-handed $U(1)$ interactions leave visible final decay products without produing neutrinos.
For a benchmark of the model parameters consistent with the current phenomenological constraints,
 we find that the travel distance of the LLP can be $\mathcal{O}(100)$~m and the LLP production
 cross section at the 14 TeV LHC can be $\mathcal{O}(10)$ fb. 
Thus, we conclude that the LLP can be discovered at the MATHUSLA with a sufficiently large
 number of LLP decay events inside the MATHUSLA detector.
\end{abstract}



 
\maketitle

\section{Introduction}

Explanation of the neutrino oscillation phenomena between neutrino flavor species
due to nonzero neutrino mass and the existence of dark matter (DM) in the Universe require
 undiscovered particles and interactions of beyond the standard model (BSM) of particle physics. 
Nevertheless, any evidence of new particles in BSM have
 not yet been reported in any ongoing experiments.
A reason of nondiscovery of a new particle might be due to not their large mass but
 their extremely weak interaction strength; in other words, the relevant coupling constant is very small. 
Even if coupling constants and hence the resultant production cross sections
of new particles are very small,
 we may expect a certain amount of production of new particles if experiments
 are conducted with huge luminosity. 
The small coupling constant makes the new particle long-lived, and signals of such long-lived particles (LLPs), if they are electrically neutral, would be discovered through their ``displaced vertex'' signatures. 
In fact, displaced vertex signatures have been sought at,
 for instance, ATLAS~\cite{ATLAS:2022pib} and FASER~\cite{FASER:2023tle} in the Large Hadron Collider (LHC).
Some near-future experiments such as FASER2~\cite{FASER:2018eoc} will start soon, and
 several other future experiments have been proposed, targeting various range of LLP masses and decay lengths. 
Among such proposed future experiments for the displaced vertex search, 
 MAssive Timing Hodoscope for Ultra Stable neutraL pArticles (MATHUSLA)
 experiment is striking, because it will be able to search for LLPs
 with a decay length of the order of $100$ m~\cite{Chou:2016lxi}. 
Theoretical studies on the search sensitivity for various BSM models have been examined in Refs.~\cite{BhupalDev:2016nfr,Evans:2017lvd,Helo:2018qej,Deppisch:2018eth,Jana:2018rdf,Bauer:2018uxu,Curtin:2018ees,Berlin:2018jbm,Dercks:2018eua,Deppisch:2019kvs,No:2019gvl,Wang:2019xvx,Jodlowski:2019ycu,Bolton:2019pcu,Hirsch:2020klk,Jana:2020qzn,Gehrlein:2021hsk,Sen:2021fha,Guo:2021vpb,Bhattacherjee:2021rml,Du:2021cmt,Kamada:2021cow,Bertuzzo:2022ozu,Liu:2022ugx,Bandyopadhyay:2022mej,Mao:2023zzk,Jodlowski:2023yne,Fitzpatrick:2023xks,Curtin:2023skh,Batz:2023zef,Deppisch:2023sga,Bernal:2023coo,Bishara:2024rtp,Liu:2024azc,deVries:2024mla,Liebersbach:2024kzc}.

The null results in dark matter direct detection experiments constrain models of
 Weakly Interacting Massive Particles (WIMPs) DM~\cite{XENON:2023cxc}.
The null results in indirect searches of DM also constrain the present DM annihilation cross section
 and exclude WIMPs annihilating into $b\bar{b}$ quarks in s-wave processes with a cross section of
 about $3 \times 10^{-26}$ cm${}^3/$s, which is the typical value for thermal WIMPs,
 with masses up to a few hundreds of GeV~\cite{Cuoco:2016eej,McDaniel:2023bju}.
Despite such stringent constraints, there are still many viable WIMP models.
One of those model proposed in Ref.~\cite{Okada:2019sbb} is
 an extra $U(1)$ interacting scalar inelastic dark matter, which is consistent
 with null results of direct DM searches due to
 its inelastic nature~\cite{Hall:1997ah,TuckerSmith:2001hy} and
 for indirect DM search because thermal abundance in the early Universe is
 fixed by the coannihilation cross section~\cite{Griest:1990kh,Edsjo:1997bg}
 rather than the self-annihilation cross section~\cite{KolbTurner}.

In the models of Ref.~\cite{Okada:2019sbb}, both the DM state and the heavier state
 relevant for both coannihilation and inelastic scattering of the DM state originate from
 one complex scalar field. 
Since they belong to a single complex scalar field,
 those two states are naturally well degenerate. 
The decay rate of the heavier state into the lighter one (DM state)
 would be very small due to its very small phase space volume, so
 that the heavier state could be a natural candidate for the LLP that we are interested in. 
To examine the MATHUSLA prospects, inelastic DM models have been proposed~\cite{Berlin:2018jbm,Guo:2021vpb,Bertuzzo:2022ozu}\footnote{The mass splitting in Refs.~\cite{Guo:2021vpb,Bertuzzo:2022ozu} is generated by an explicit gauge symmetry breaking term, and models have not been formulated in gauge invariant manner.}. While the longevity of LLP in fermionic inelastic DM model proposed in Ref.~\cite{Berlin:2018jbm}
 is due to the mass degeneracy as well as its tiny kinetic mixing of dark photons,
 the LLP in our scalar inelastic DM model is long-lived due to only about $1 \%$ mass degeneracy.

This paper is organized as follows:  
In Section \ref{Sec:Models}, we describe the extra gauge $U(1)$ interacting inelastic model,
 which is a variant of the original models in Ref.~\cite{Okada:2019sbb}. 
We identify a parameter set to reproduce thermal DM abundance and a benchmark point in Section~\ref{Sec:DM}.
In Section~\ref{Sec:LLP}, for the benchmark point,
 we present the search prospects of LLPs by the MATHUSLA experiment. 
Section~\ref{Sec:Summary} is devoted to our summary.

\section{The Model}
\label{Sec:Models}

There are several possibilities of the anomaly-free gauged $U(1)$ extension of the Standard Model.
Among them, the best studied model is the flavor
 universal $U(1)_{B-L}$~\cite{Pati:1973uk,Davidson:1978pm,Mohapatra:1980qe,Mohapatra:1980}.
However, because the new neutral gauge boson $Z'$ in the $U(1)_{B-L}$ model interacts
 with the first and the second generation of SM fermions,
 the experimental constraints on the couplings are very stringent~\cite{Carena:2004xs,Amrith:2018yfb,Das:2019fee}.
Thus, we may not expect a sizable production cross section for the LLP production from the $Z'$ boson decay.
Note that anomaly cancellation of the $U(1)_{B-L}$ model is realized in each generation of fermions;
 hence it is generally possible for only a particular generation to be charged under $U(1)$.
A model where only the third-generation fermions are charged, $U(1)_{(B-L)_3}$,
 is such a choice~\cite{Okada:2019sbb,Babu:2017olk,Alonso:2017uky,Bian:2017rpg,Cox:2017rgn}
 and has been well studied. 
As expected, the experimental constraints on the $U(1)_{(B-L)_3}$ model are much weaker
 compared with the universal $U(1)_{B-L}$
 model~\cite{delAmoSanchez:2010bt,Faroughy:2016osc,Aaboud:2017sjh,Chun:2018ibr,Elahi:2019drj}.
It is easy to generalize the $U(1)_{B-L}$ model to the $U(1)_X$ model by assigning the $U(1)$ charge for a field as a linear combination of its $U(1)_Y$ and $U(1)_{B-L}$ charges. 
Since $U(1)_Y$ and $U(1)_{B-L}$ are independently anomaly free, the $U(1)_X$ is automatically anomaly free~\cite{Appelquist:2002mw,Oda:2015gna,Das:2016zue}.
A special case is the so-called $U(1)_R$ model, where only right-handed (RH) SM fermions and RH neutrinos are charged under $U(1)$~\cite{Jung:2009jz}.
In this paper, we generalize $U(1)_{(B-L)_3}$ to $U(1)_{X_3}$, 
 where $X_3$ denotes a linear combination of $U(1)_{Y_3}$ and $U(1)_{(B-L)_3}$.
We introduce an $U(1)_{X_3}$ charged scalar field $\phi_1$ with its charge $+1$
 in addition to the minimal particle contents.
The total particle contents are listed in Table~\ref{table:charge}.
The field $\phi_2$ is responsible for breaking the $U(1)_{X_3}$ gauge symmetry and
 $x_H$ is a real parameter that parameterizes a combination weight of $(B-L)_3$ and $Y_3$.
Here, we note that the $U(1)_{(B-L)_3}$ and $U(1)_{X_3}$ models with a minimal Higgs doublet
 cannot reproduce realistic fermion masses and flavor mixings due to the gauge symmetry.
Since the details of those UV completions are irrelevant for our main discussion
 on DM and LLP phenomenology, we work on the simplified particle content in Table~\ref{table:charge}.
Nevertheless, we note an example of successful UV completions of $U(1)_{R_3}$ in Appendix~\ref{UV example}\footnote{
 A few successful UV completions for $U(1)_{(B-L)_3}$ can be found in Refs.~\cite{Babu:2017olk,Alonso:2017uky}.}.

\begin{table}[t]
\begin{center}
\begin{tabular}{|c|ccc|c|}
\hline
            &  SU(3)$_c$ & SU(2)$_L$ & U(1)$_Y$ & U(1)$_{X_3}$  \\ 
\hline
$Q^{i}$     & {\bf 3 }   &  {\bf 2}  & $ 1/6$   & $\left(\frac{1}{6}x_H+\frac{1}{3}\right) \delta_{i3}$    \\
$u^{i}_{R}$ & {\bf 3 }   &  {\bf 1}  & $ 2/3$   & $\left(\frac{2}{3}x_H+\frac{1}{3}\right) \delta_{i3}$    \\
$d^{i}_{R}$ & {\bf 3 }   &  {\bf 1}  & $-1/3$   & $\left(-\frac{1}{3}x_H+\frac{1}{3}\right) \delta_{i3}$   \\
\hline
$L^{i}$     & {\bf 1 }   &  {\bf 2}  & $-1/2$   & $\left(-\frac{1}{2}x_H-1\right)\delta_{i3} $     \\
$e^{i}_{R}$ & {\bf 1 }   &  {\bf 1}  & $-1$     & $\left(-x_H-1\right)\delta_{i3}$     \\
\hline
$\Phi$      & {\bf 1 }   &  {\bf 2}  & $ 1/2$   & $0 $   \\  
\hline
$N^{i}_{R}$ & {\bf 1 }   &  {\bf 1}  & $0$      & $-\delta_{i3}$      \\
$\phi_1$    & {\bf 1 }   &  {\bf 1}  & $ 0$     & $ + 1 $  \\
$\phi_2$    & {\bf 1 }   &  {\bf 1}  & $ 0$     & $ + 2 $  \\ 
\hline
\end{tabular}
\end{center}
\caption{
The particle contents of our $U(1)_{X_3}$ model. 
In addition to the SM particle content ($i=1,2,3$), three RH neutrinos  
  ($N_R^i$ ($i=1, 2, 3$)) and two $U(1)_{X_3}$ scalar fields ($\phi_1$ and $\phi_2$) are introduced.   
 The scalar $\phi_2$ is the Higgs field and develops its VEV to break the $U(1)_{X_3}$ symmetry, while $\phi_1$ has no VEV and includes the inelastic DM and its partner LLP as its components.
}
\label{table:charge}
\end{table}

The $U(1)_{X_3}$ gauge interaction for an SM chiral fermion and RH neutrinos ($f_{L/R}$) can be read from the usual covariant derivative, 
\begin{align}
\mathcal{L}_\mathrm{int} = \sum_{f_j} \overline{f_j}\gamma^{\mu} X_{\mu}g_{X_3}\left(q_{f_{jL}} P_L+q_{f_{jR}} P_R\right)f_j ,
\end{align}
 where $X^{\mu}$ is the $U(1)_{X_3}$ gauge field, $g_{X_3}$ is the gauge coupling constant, and $q_{f_{j L/R}}$ is a $U(1)_{X_3}$ charge of $f_{j L/R}$ (see Table~\ref{table:charge}). Here, we do not consider the gauge kinetic mixing because the mixing between $Z$ and $Z'$ bosons is experimentally severely constrained;  in fact the mixing can also be canceled in a UV model (see Appendix~\ref{UV example}.).
According to the experimental constraints, we have chosen our renormalization condition
 so as to vanish the kinetic mixing at the scale of $P$ mass.

The scalar potential, which is gauge invariant and renormalizable, is expressed as~\cite{Okada:2018xdh,Chao:2017ilw}
\begin{align}
V(\Phi, \phi_1, \phi_2 )
 =& - M^2_{\Phi} |\Phi|^2 + \frac{\lambda}{2} |\Phi|^4 + M^2_{\phi_1} \phi_1\phi_1^{\dagger} - M^2_{\phi_2} \phi_2\phi_2^{\dagger}   \nonumber \\
 & + \frac{1}{2} \lambda_1 (\phi_1 \phi_1^{\dagger})^2+\frac{1}{2}\lambda_2 (\phi_2\phi_2^{\dagger} )^2
 +\lambda_3 \phi_1\phi_1^{\dagger} (\phi_2 \phi_2^{\dagger}) \nonumber \\
 & +  (\lambda_4 \phi_1\phi_1^{\dagger} + \lambda_5 \phi_2\phi_2^{\dagger})|\Phi|^2
 - A (\phi_1 \phi_1 \phi_2^{\dagger} + \phi_1^{\dagger} \phi_1^{\dagger} \phi_2 ) ,
\label{eq:totalpotential}
\end{align}
with $\Phi$ being the SM Higgs doublet field.
All parameters, $M^2_i, \lambda_i$ and $A$, in the potential~(\ref{eq:totalpotential}) are taken to be real and positive.

\subsection{Dark matter mass and interactions}

At $U(1)_{X_3}$ and the electroweak (EW) symmetry breaking vacuum, the SM Higgs field and
 the $U(1)$ Higgs field are expanded around the VEVs $v$ and $v_2$ as (in the unitary gauge)
\begin{align}
\Phi =& \left( \begin{array}{c}
          0 \\
          \frac{v + \varphi}{\sqrt{2}} \\
         \end{array}
  \right), \\
\phi_1 =& \frac{ S + i P }{\sqrt{2}} ,\\
\phi_2 =& \frac{ v_2 + \varphi_2 }{\sqrt{2}} .
\end{align}
The physical states ($\varphi$ and $\varphi_2$) are diagonalized 
 to the mass eigenstates ($h$ and $H$) with masses $m_h$ and $m_H$ as 
 \begin{equation}
  \left( \begin{array}{c}
          \varphi  \\
          \varphi_2  \\
         \end{array}\right)
   = 
  \left( \begin{array}{cc}
          \cos\alpha & \sin\alpha\\
          -\sin\alpha & \cos\alpha\\
         \end{array}\right)
  \left( \begin{array}{c}
          h \\
          H \\
         \end{array}\right) .
 \end{equation}
For a small mixing angle $\alpha$, $h$ is identified with the SM-like Higgs boson.
Hereafter, we set $\alpha$ negligibly small.
With $U(1)_{X_3}$ and the EW symmetry breaking, the $Z^\prime$ boson (the mass eigenstate after the $X$ becomes massive), $S$ and $P$ acquire their masses, respectively, as 
\begin{align}
  m_{Z^\prime}^2 =& g_{X_3}^2 4v_2^2, \\
  m_S^2 =& M_{\phi_1}^2+\frac{1}{2}\lambda_3v_2^2+\frac{1}{2}\lambda_4 v^2-\sqrt{2}A v_2 , \label{eq:Smass}\\
  m_P^2 =& M_{\phi_1}^2+\frac{1}{2}\lambda_3v_2^2+\frac{1}{2}\lambda_4 v^2+\sqrt{2}A v_2 . \label{eq:Pmass}
\end{align}
Note that the parameter $A$ controls the mass splitting between $S$ and $P$. 
Since we take $A$ positive, $S$ is lighter than $P$ and becomes the DM candidate.
The other choice of $A<0$ causes no essential difference in our final results, except that $P$ is the DM particle in the case.

Gauge interaction of the DM particle is expressed as
\begin{equation}
\mathcal{L}_\mathrm{int} = g_{X_3} Z'{}^{\mu}\left( (\partial_{\mu}S) P- S \partial_{\mu}P \right) ,
\label{eq:Lag:gauge-DM-DM}
\end{equation}
 and similarly third-generation quarks and leptons also interact with the $Z'$ boson with
 corresponding charges.
The absence of $Z'$-DM-DM coupling indicates that the $Z'$-mediating DM scattering
 with a nucleon is inelastic and ineffective for a mass splitting larger than
 the maximal energy transfer in the scatterings~\cite{Hall:1997ah,TuckerSmith:2001hy}.
Elastic scattering through Higgs bosons exchange~\cite{Jungman:1995df} can
 be negligible for a very small Higgs mixing $\alpha$ and the coupling $\lambda_4$.
We also set the coupling constant $A$ very small, so that $S$ and $P$ are well degenerate in mass.
This degeneracy is crucial not only for the scalar $P$ being long-lived,
 but also for reproducing the observed DM relic density through $S$ and $P$ coannihilation process.

\subsection{The decay width of $Z'$ boson}
\label{Sec:Z'width}

\begin{figure}[htbp]
\centering
\includegraphics[clip,width=11.0cm]{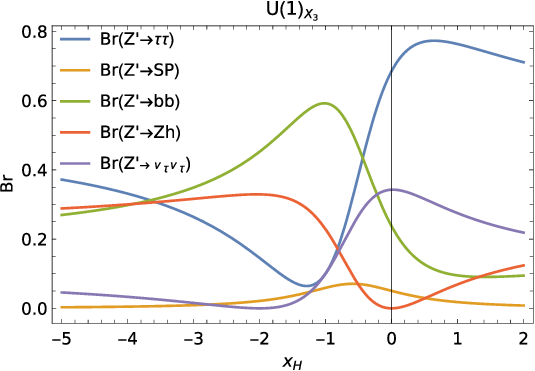}
\caption{
Branching ratios of the $Z'$ decay for $g_{X}=0.1, m_S=130$ GeV and $m_{Z'}=350$ GeV. }
\label{Fig:Br-xH}
\end{figure}

The partial decay width of $Z' \rightarrow f\bar{f}$ is given by
\begin{align}
 \Gamma(Z' \rightarrow f\bar{f}) = & \frac{1}{48\pi m_{Z'}^2}N_c|\mathcal{M}|^2\sqrt{m_{Z'}^2-4 m_f^2},  \label{eq:Z'decaytoff} \\
 |\mathcal{M}|^2 =& 2 g_{X_3}^2 \left(q_{fL}^2 \left(m_{Z'}^2-m_f^2\right)+6 q_{fL} q_{fR} m_f^2+q_{fR}^2 \left(m_{Z'}^2-m_f^2\right)\right) \label{eq:calM-Z'toff},
\end{align}
 where the color number $N_c$ is $3$ for quarks and $1$ for leptons.
If $m_f \ll m_{Z'}$, then Eq.~(\ref{eq:Z'decaytoff}) is reduced to  
\begin{align}
 \Gamma(Z' \rightarrow f\bar{f}) \simeq & \frac{g_{X_3}^2 N_c m_{Z'} }{24\pi} \left(q_{fL}^2+q_{fR}^2\right) .
\end{align}
We obtain the partial decay width of $Z'$ into $S$ and $P$ as
\begin{align}
\Gamma(Z'\rightarrow SP) = \frac{g_{X_3}^2}{48 \pi} \frac{(m_{Z'}^2-(m_P-m_S)^2)^{3/2}(m_{Z'}^2-(m_P+m_S)^2)^{3/2} }{m_{Z'}^5} ,
\end{align}
 from the vertex (\ref{eq:Lag:gauge-DM-DM}).
The decay branching ratios are shown in Fig.~\ref{Fig:Br-xH}.
Note that $Br(Z'\rightarrow \nu_{\tau}\nu_{\tau})$ vanishes at $x_H=-2$, which corresponds to right-handed $U(1)$, $U(1)_R$. 
This fact plays an important role in the following discussion.

\subsection{Decay of $P$}

\begin{figure}[htbp]
\centering
\includegraphics[clip,width=11.0cm]{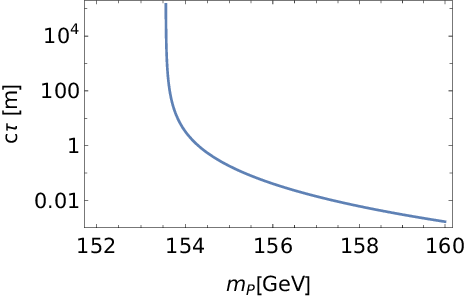}
\caption{
$c\tau$ for $g_{R_3}=0.1, m_S=150$ GeV and $m_{Z'}=350$ GeV.
Near the threshold $m_P-m_S \simeq 2m_{\tau}$ with $m_{\tau}$ being the tau-lepton mass, $c\tau$ is very long. }
\label{Fig:ctau}
\end{figure}

If $S$ and $P$ are strongly degenerate so that the mass difference is much smaller than the mass of $Z'$,
the two-body decay of $P$ is kinematically forbidden. Thus, for the main decay mode, 
$P \rightarrow S Z'{}^* \rightarrow Sf \bar{f}$, the total decay width ($\Gamma$), or equivalently the inverse of the lifetime ($\tau$), 
\begin{align}
\Gamma = \frac{1}{\tau} =\frac{1}{2p^0}\int\overline{|\mathcal{M} (P\rightarrow S f\bar{f})|^2}dQ ,
\end{align}
 is suppressed by the phase space volume~\cite{Mohapatra:2023aei}.
As a result, $P$ is short-lived in cosmology but can be long-lived in collider experiments.
The expected travel distance $c\tau$ for $x_H=-2$, in other words $U(1)_{R_3}$,
 is shown as the function of its mass $m_P$ in Fig.~\ref{Fig:ctau}.
We note that the mass difference, 
\begin{align}
 m_P^2-m_S^2 = 2\sqrt{2}A v_2,
\end{align}
 from Eqs.~(\ref{eq:Smass}) and (\ref{eq:Pmass}), is always adjustable by suitably
 choosing the free parameter $A$ such that $c\tau$ becomes about $100$ m, which is ideal for the detection of $P$ decay inside the MATHUSLA detector.
Here and hereafter, we have forcused on the $U(1)_{R_3}$ case.
For $x_H \neq -2$, the decay mode of $P\rightarrow SZ'{}^* \rightarrow S \nu_{\tau}\nu_{\tau}$ exits and becomes the dominant mode for $M_P-m_S \leq 2 m_{\tau}$.
In this case, the LPP $P$ provides only invisible decay products.

\section{Dark matter and LHC constraints}
\label{Sec:DM}

For the evaluation of prospects at MATHUSLA in the next section, in this section,
 we will find a benchmark point thst satisfies the thermal DM abundance and the latest LHC constraints.
As metioned above, we concentrate on the $U(1)_{R_3}$ model.
The LLP $P$ in other general $U(1)_{X_3}~(x_H\neq -2)$ may decay into neutrinos
 which are invisible for the MATHUSLA detector.
Thus, the $U(1)_{R_3}$ model is the best choice from the viewpoint of the detection with a large decay length of $\mathcal{O}(100)$ m.

\subsection{Thermal relic abundance}

\begin{figure}[htbp]
\centering
\includegraphics[width=8.0cm]{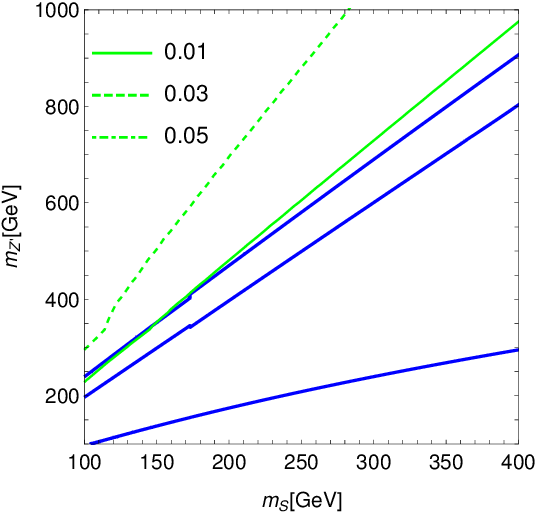}
\includegraphics[width=8.0cm]{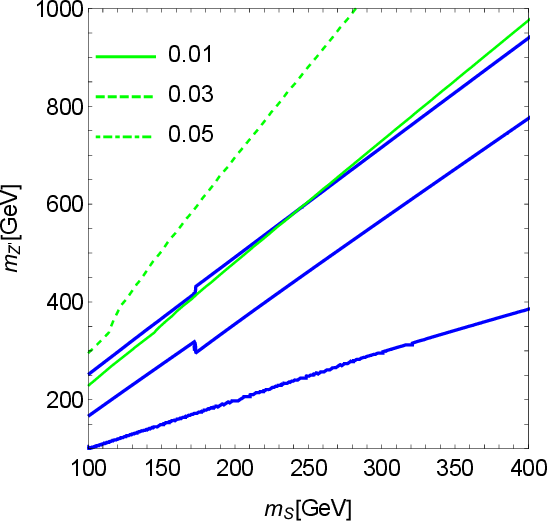}
\caption{
The contour (in blue) along which the observed DM relic abundance $\Omega h^2 \simeq 0.1$ is reproduced for $g_{R_3}=0.1$ and $x_H=-2$. Green curves are contours of $\mathrm{Br}(Z' \rightarrow SP)$. Here, $g_{R_3}=0.1 $ in the left pannel and $=0.2$ in the right pannel.
}
\label{Fig:r3thermal03}
\end{figure}

We estimate the thermal relic abundance of the real scalar DM, $S$, by solving the Boltzmann equation,
\begin{equation}
 \frac{d n }{dt}+3H n =-\langle\sigma_\mathrm{eff} v\rangle ( n^2 - n_{\rm EQ}^2),
\label{eq:boltzman}
\end{equation}
 where $H$ and $n_{\rm EQ}$ are the Hubble parameter and the DM number density at thermal equilibrium,
 respectively~\cite{KolbTurner}.  
In our model, the main annihilation mode is coannihilation $S P \rightarrow f\bar{f}$ through $s$-channel $Z'$ exchange for $m_{Z'} > m_S$ and the annihilation mode $S S \rightarrow Z^\prime Z'$ by $u(t)$-channel $P$ exchange for $m_{Z'} < m_S$~\cite{Okada:2019sbb}.
We use the effective thermal averaged annihilation cross section
\begin{equation}
 \langle\sigma_\mathrm{eff}v\rangle  = \sum_{i,j= S,P} \langle\sigma_{ij} v_{ij}\rangle\frac{n_i}{n_{\rm EQ}}\frac{n_j}{n_{\rm EQ}},
\end{equation}
 to include the coannihilation effects properly and $n$ in Eq.~(\ref{eq:boltzman}) should be understood as $n = \sum_i n_i$ for $i = S, P$~\cite{Griest:1990kh,Edsjo:1997bg}.
The contours of thermal DM abundance $\Omega h^2 \simeq 0.1$~\cite{Planck:2018vyg} (blue curve) and $\mathrm{Br}(Z' \rightarrow SP)$ (green curves) for the $U(1)_{R_3}$ model are shown in Fig.~\ref{Fig:r3thermal03}.

\subsection{Benchmark points}

To find a viable benchmark point, we take the $Z'$ boson search at the LHC into account.
With the decay rates of $Z'$ estimated in Section~\ref{Sec:Z'width},
 since the total $Z'$ boson decay width is very narrow, as in Ref.~\cite{Das:2019fee},
 we use the narrow width approximation to evaluate the $Z'$ boson production cross section 
\begin{align}
\sigma(pp \rightarrow X) &= 2\sum_{q,\bar{q}}\int dx\int dy f_q(x,Q) f_{\bar{q}}(x,Q)\hat{\sigma}(\hat{s}) , \label{Eq:sigma_ppX}\\
\hat{\sigma}(\hat{s}) &= \frac{4\pi^2}{3}\frac{\Gamma_X(X \rightarrow q\bar{q})}{m_X}\delta(\hat{s}-m_X^2) , 
\end{align}
where $f_q$ and $f_{\bar{q}}$ are the parton distribution function (PDF) for a quark and antiquark ($b$ and $\bar{b}$ in our model), $\hat{s}=x y s$ is the invariant mass squared of colliding quarks for the center of mass energy $s$.
The factor $2$ in Eq.~(\ref{Eq:sigma_ppX}) accounts for two ways of $q$ coming from which proton out of two colliding protons.
We employ PDFs of CTEQ6L~\cite{Pumplin:2002vw} with a factorization scale $Q = m_{Z'}$ for simplicity.
 
Since the severest bound on the $Z'$ boson is from the dilepton channel ($\ell=\tau$)~\cite{CMS:2024pjt,ATLAS:2019erb}\footnote{
We note that the present experimental bound from the $Z' \rightarrow Z h$ mode reported by the ATLAS~\cite{ATLAS:2022enb} and CMS~\cite{CMS:2024phk} Collaborations for $m_{Z'} = \mathcal{O}(100)$ GeV are about one order of magnitude weaker
than that from the $\tau^+ \tau^-$ mode.}, we evaluate the production cross section for the process $pp\rightarrow \tau^+\tau^-$ given by
\begin{align}
\sigma(pp \rightarrow Z')\mathrm{BR}(Z'\rightarrow \tau^+\tau^-),
\end{align}
 where the branching ratio is calculated from Eq.~(\ref{eq:Z'decaytoff}), 
and compare it with the CMS results~\cite{CMS:2024pjt}\footnote{The ATLAS bound~\cite{ATLAS:2019erb} is somewhat weaker than that from the CMS.}.
The resultant production cross section is shown in the second column in Table~\ref{table:Xsection}.
The production cross section is consistent with and smaller than
 the latest most stringent bound from the CMS~\cite{CMS:2024pjt} at
 the LHC.
We summarize parameters and masses of relevant particles of benchmark points with different values
of the $U(1)_{R_3}$ gauge coupling constant in Table~\ref{table:points}.
\begin{table}[htb]
\centering
\begin{tabular}{|c|c|c|c|c|c|}
\hline
            && $x_H$  & $g_{X_3}$  & $m_{Z'}$ & $m_S$    \\ 
\hline
\hline
$BP_{1}$   && $-2$   & $0.1$  & $351$ GeV & $150$ GeV  \\
\hline
$BP_{2}$   && $-2$   & $0.2$  & $850$ GeV & $351$ GeV  \\
\hline
\end{tabular}
\caption{The values of parameters and the masses of $Z'$ and $S$ in benchmark points.
}
\label{table:points}
\end{table}

\section{Prospect for MATHUSLA}
\label{Sec:LLP}

The travel distance $c\tau \simeq 100$ m is the most sensitive range at MATHUSLA~
\cite{Chou:2016lxi,Curtin:2017izq,Curtin:2017bxr}.
The pair production cross section of such an LLP $\chi$ for its $5\sigma$ discovery
 can be read as $\sigma_{\chi\chi} > 0.3$ fb from Fig.~1 in Ref.~\cite{Jana:2018rdf}.
On the other hand, in our scenario, the LLP $P$ is not pair produced but singly produced
 through $pp \rightarrow Z' \rightarrow SP$.
Thus, the required production cross section for the LLP $P$ to be discovered
 at MATHUSLA would be set to be larger than $0.6$ fb.
In Table~\ref{table:Xsection}, we list the $SP$ production cross section at $13$ TeV LHC and $14$ TeV LHC for our benchmark on Table~\ref{table:points} satisfying the DM abundance and the LHC bounds.
We find that MATHUSLA will be able to discover those LLPs
 because the production cross section $\sigma_{SP}$ is sufficiently larger than $0.6$ fb.

\begin{table}[ht]
\begin{center}
\begin{tabular}{|c|c|c|c|c|c|c|}
\hline
            &&  $\sigma(pp \rightarrow Z' \rightarrow \tau\tau)$ (CMS bound on it) && $\sigma_{SP}$ @$13$TeV LHC & $\sigma_{SP}$ @$14$TeV LHC  \\ 
\hline
$BP_{1}$   && $154$ fb ($<203$ fb)  && $10.8$ fb  & $12.9$ fb  \\
\hline
$BP_{2}$   && $7.78$ fb ($<8.24$ fb)  && $0.697$ fb  & $0.888$ fb  \\
\hline
\end{tabular}
\end{center}
\caption{Estimated production cross section for tau pairs and $SP$ pairs in benchmark points.}
\label{table:Xsection}
\end{table}

\section{Summary}
\label{Sec:Summary}

We have proposed a simple extension of the SM with an extra $U(1)_{R_3}$ gauge interaction
 with a scalar particle of DM candidate with the charge $1$, third generation of SM fermions
 and third generation of right-handed neutrinos.
After the SM singlet $U(1)_{R_3}$ breaking scalar with charge $2$ develops a VEV,
 the gauge boson acquires a mass, 
 and a tiny mass splitting between the real and imaginary components of the charge $1$ scalar appears 
 through the scalar trilinear interaction.
The lighter scalar $S$ is an inelastic DM candidate with a heavier state $P$.
Due to the mass degeneracy, the slightly heavier state $P$ is long-lived and
 will be able to be discovered as LLPs.

We have calculated the production cross section of LLP $P$
 through the process $pp\rightarrow Z' \rightarrow SP$ at MATHUSLA for a benchmark point 
 that satisfies stringent LHC bounds and thermal DM abundance. 
The production cross section of the benchmark point turns out to be about $10$ fb, which is about an order of magnitude larger than the cross section required by $5\sigma$ discovery as a LLP at the MATHUSLA.
Thus, we conclude that the heavier state of inelastic DM in our model can be discovered at MATHUSLA.


\section*{Acknowledgments}
This work is supported in part by the U.S. DOE Grant No.~DE-SC0012447 and DE-SC0023713 (N.O.) and
KAKENHI Grants No.~JP23K03402 (O.S.).

\newpage
\appendix

\section{An extension of Higgs sector}
\label{UV example}

We note a possible UV completed model to generate top qurak mass and
 how gauge kinetic mixing parameter affects the mixing between $Z$ and $Z'$.
We may consider an extension with three Higgs doublet fields $\Phi_1, \Phi_2$ and $\Phi_3$ and
some singlet scalar fields to eliminate unwanted NG modes. 
The charge assignments of those Higgs fields are in Tab.~\ref{table:extended:charge}.
Those doublet fields develop the VEVs as
\begin{align}
\Phi_i =& \left( \begin{array}{c}
          0 \\
          \frac{v_i}{\sqrt{2}} \\
         \end{array}
  \right) .
\end{align}
The quark Yukawa interactions are expressed as
\begin{align}
 \mathcal{L}_\mathrm{Yukawa} = 
  & \sum_{i,j}^{1,2}\left(-y_{ij}^{u} \overline{Q}_i \tilde{\Phi}_1 u_{jR} -y_{ij}^{d} \overline{Q}_i \Phi_1 d_{jR}-y_{3j}^{u} \overline{Q}_3 \tilde{\Phi}_1 u_{jR} -y_{3j}^{d} \overline{Q}_3 \Phi_1 d_{jR}\right) \nonumber \\
  & -y_{i3}^{u} \overline{Q}_i \tilde{\Phi}_2 u_{3R} -y_{i3}^{d} \overline{Q}_i \Phi_2 d_{3R} -y_{33}^{u} \overline{Q}_3 \tilde{\Phi}_2 u_{3R} -y_{33}^{d} \overline{Q}_3 \Phi_2 d_{3R} + \mathrm{ H.c.} ,
\label{Lag:quarkyukawa} 
\end{align}
and there are non-vanishing entries for all $3$ by $3$ elements. The mass of top quark is generated through the VEV $v_2$.
At the same time, the mass matrix of neutral gauge bosons $\hat{Z}$ and $\hat{X}$ in the interaction basis are given by
\begin{equation}
\left( \begin{array}{cc}
     M_{\hat{Z}\hat{Z}}^2 & M_{\hat{Z}\hat{X}}^2\\
     M_{\hat{X}\hat{Z}}^2 & M_{\hat{X}\hat{X}}^2\\
       \end{array}\right) =
\left( \begin{array}{cc}
     \frac{g_2^2 v^2}{8c_W^2} & -\frac{g_2 g_X}{4 c_W}\left( -v_2^2+q_{X_3} v_3^2 \right) \\
     -\frac{g_2 g_X }{4 c_W}\left( -v_2^2+q_{X_3} v_3^2 \right) &
     \frac{ g_X^2}{2} \left( v_2^2+q_{X_3}^2 v_3^2+4 v_{\phi}^2\right) \\
       \end{array}\right)  ,
\end{equation}
 with $v^2=v_1^2+v_2^2+v_3^2$, where $g_2$ is the $SU(2)_L$ gauge coupling, $g_X$ is the $U(1)_{R_3}$ gauge coupling,
$v_{\phi}$ is the VEV of singlet Higgs field with the $U(1)_{R_3}$ charge $-2$,
 the the $U(1)_{R_3}$ charges of $\Phi_1, \Phi_2$ and $\Phi_3$ are taken to be $0, -1, q_{X_3}$, respectively and
 $c_W$ is the cosine of the Weinberg angle.
 Note that the mass mixing between $\hat{Z}$ and $\hat{X}$ is tiny for $-v_2^2+q_{X_3} v_3^2 \simeq 0$~\cite{Seto:2020jal}.
The field redefinition by the orthogonal matrix resolve the gauge kinetic mixing 
\begin{align}
\mathcal{L}_{\mathrm{Gauge}} \supset  \frac{\sin\epsilon}{2}\hat{B}_{\mu\nu}\hat{X}^{\mu\nu},
\label{Lag:gauge} 
\end{align}
but induces additional mass mixings. The resultant mass matrix for the canonical kinetic terms becomes
\begin{equation}
\left(
\begin{array}{cc}
  M^2_{\hat{Z}\hat{Z}} & \frac{M^2_{\hat{Z}\hat{X}}}{c_{\epsilon}}-M^2_{\hat{Z}\hat{Z}} s_W t_{\epsilon} \\
  \frac{M^2_{\hat{Z}\hat{X}}}{c_{\epsilon}}-M^2_{\hat{Z}\hat{Z}} s_W t_{\epsilon} &  \frac{s_W c_{\epsilon}t_{\epsilon}}{c_{\epsilon}^2}  \left(  M^2_{\hat{Z}\hat{Z}}s_W  c_{\epsilon} t_{\epsilon}-2 M^2_{\hat{Z}\hat{X}} \right)+\frac{1}{c_{\epsilon}^2} M^2_{\hat{X}\hat{X}} \\
\end{array}
\right) .
\end{equation}
The magnitude of the off-diagonal component, or equvalently the $Z-Z'$ mixing, is by experiments 
stringently constrained as
 $ M^2_{\hat{Z}\hat{X}}-M^2_{\hat{Z}\hat{Z}} s_W s_{\epsilon}\lesssim (0.1 \,\mathrm{GeV})^2$~\cite{ParticleDataGroup:2024cfk},
 so that the shift of the $Z$ boson mass due to the mixing must be within the error of its measurement.
In our model, by seting the ratio $v_2/v_3$ in
 $M^2_{\hat{Z}\hat{X}}$ appropriately so that the off-diagonal element can be as small as to be consistent, 
the mixing between $Z$ and $Z'$ is always negligible even under non-vanishing gauge kinetic parameter $\epsilon$.

One can easily see that physics of dark matter and $Z'$ described in the main text are intact 
for the above extension.

\begin{table}[h]
\begin{center}
\begin{tabular}{|c|ccc|c|c|}
\hline
            &  SU(3)$_c$ & SU(2)$_L$ & U(1)$_Y$ & U(1)$_{R_3}$ & $Z_2$ \\ 
\hline
$Q^{i}$     & {\bf 3 }   &  {\bf 2}  & $ 1/6$   & $0$  & $+$   \\
$u^{i}_{R}$ & {\bf 3 }   &  {\bf 1}  & $ 2/3$   & $ -\delta_{i3}$   & $+$   \\
$d^{i}_{R}$ & {\bf 3 }   &  {\bf 1}  & $-1/3$   & $ \delta_{i3}$  & $+$   \\
\hline
$\Phi_1$      & {\bf 1 }   &  {\bf 2}  & $ 1/2$   & $0 $   & $+$  \\  
$\Phi_2$      & {\bf 1 }   &  {\bf 2}  & $ 1/2$   & $-1  $  & $+$   \\  
$\Phi_3$      & {\bf 1 }   &  {\bf 2}  & $ 1/2$   & $q_{X3}  $   & $+$  \\  
\hline
$\phi_1$    & {\bf 1 }   &  {\bf 1}  & $ 0$     & $ + 1 $   & $-$\\
$\phi_1'$   & {\bf 1 }   &  {\bf 1}  & $ 0$    & $ + 1 $   & $+$ \\ 
$\phi_2$    & {\bf 1 }   &  {\bf 1}  & $ 0$     & $ + 2 $   & $+$ \\ 
$\phi_3$    & {\bf 1 }   &  {\bf 1}  & $ 0$     & $q_{X3}$   & $+$ \\ 
\hline
\end{tabular}
\end{center}
\caption{
The short list of particle contents with an extended Higgs sector. 
$\phi_1'$ and $\phi_3$ provide the masses of NG bosons to be heavy psuedo scalars.
A $Z_2$ parity distinguishes $\phi_1$ and $\phi_1'$, and let DM stable.
}
\label{table:extended:charge}
\end{table}

\section{The decay rate of $P$}

The spin averaged squared amplitude for $P(p) \rightarrow f(q_1) \bar{f}(q_2)S(q_3)$ is given by
\begin{align}
& \overline{|\mathcal{M}|^2} \nonumber \\
 =& \frac{2 g_{R_3}^4}{m_{Z'}^4 \left(-2 \left(p \cdot q_3 \right)+m_P^2+m_S^2-m_{Z'}^2\right)^2} \left(-2 \left(-m_P^2+m_S^2+m_{Z'}^2\right)^2 \left(p\cdot q_1\right)^2  \right. \nonumber \\
 &  -2\left(m_P^2-m_S^2+m_{Z'}^2\right)^2\left(q_1 \cdot q_3 \right)^2+4\left(\left(m_P^2-m_S^2\right)^2-m_{Z'}^4\right)\left(p\cdot q_1\right)\left(q_1 \cdot q_3 \right)    \nonumber \\
 &  -2\left(-m_P^2+m_S^2+m_{Z'}^2\right)^2\left(p\cdot q_1\right)\left(p\cdot q_3\right)+2\left(m_P^2-m_S^2+m_{Z'}^2\right)^2\left(q_1\cdot q_3\right)\left(p\cdot q_3\right)   \nonumber \\
 & \left.  +m_f^2 \left(-2 \left(\left(m_P^2-m_S^2\right)^2-m_{Z'}^4\right) \left( p \cdot q_3 \right)+m_{Z'}^4 \left(m_P^2+m_S^2\right)+\left(m_P^2-m_S^2\right)^2 \left(m_P^2+m_S^2-2 m_{Z'}^2\right)\right) \right. \nonumber \\
 & \left.  +\left(p\cdot q_1 \right) \left(m_{Z'}^4 \left(m_P^2-3 m_S^2\right)+2 m_{Z'}^2 \left(m_S^4-m_P^4\right)+\left(m_P^2-m_S^2\right)^2 \left(m_P^2+m_S^2\right)\right) \right. \nonumber \\
 & \left.  +\left(q_1 \cdot q_3\right) \left( m_P^4 \left(m_S^2-2 m_{Z'}^2\right)+m_P^2 \left(m_S^4+3 m_{Z'}^4\right)-m_P^6-\left(m_S^3-m_S m_{Z'}^2\right)^2\right)\right) .
\end{align}
The integration of phase space volume is reduced to 
\begin{align}
dQ =&\frac{d^3q_1}{(2\pi)^3 2q_1^0}\frac{d^3q_2}{(2\pi)^3 2q_2^0}\frac{d^3q_3}{(2\pi)^3 2q_3^0} (2\pi)^4\delta^{(4)}\left(p-q_1-q_2-q_3\right) \nonumber \\
=& \frac{1}{(2\pi)^5}\frac{d^3q_1}{2q_1^0}\frac{d^3q_3}{2q_3^0}
 \delta\left((m_P-q_1^0-q_3^0)^2-(|\mathbf{q}_1|^2+|\mathbf{q}_3|^2)-2|\mathbf{q}_1||\mathbf{q}_3|\cos\theta_{13} -m_f^2 \right) \nonumber \\
=& \frac{1}{8(2\pi)^5}dq_1^0  dq_3^0d\Omega_1 d\varphi_{13}d\cos\theta_{13} \delta\left(\frac{(m_P-q_1^0-q_3^0)^2-|\mathbf{q}_1|^2-|\mathbf{q}_3|^2-m_f^2}{2|\mathbf{q}_1||\mathbf{q}_3|}-\cos\theta_{13}\right) .
\end{align}
Then, the integration range of energy turns out to be
\begin{align}
 q_3^\mathrm{min}(q_1^0)  <  q_3^0 < q_3^\mathrm{max}(q_1^0) ,
\label{eq:q3_int_range}
\end{align}
 with
\begin{align}
& q_3^\mathrm{min}(q_1^0) = \frac{\left(m_P-q_1^0 \right) \left(m_P^2-2 m_P q_1^0+m_S^2 \right)-\sqrt{\left(\left(q_1^0 \right)^2-m_f^2 \right) \left( \left(m_P^2-2 m_P q_1^0 -m_S^2\right)^2-4m_f^2m_S^2 \right)}}{2 \left(m_f^2+m_P \left(m_P-2 q_1^0\right)\right)} , \\
& q_3^\mathrm{max}(q_1^0) = \frac{\left(m_P-q_1^0 \right) \left(m_P^2-2 m_P q_1^0+m_S^2 \right)+\sqrt{\left(\left(q_1^0 \right)^2-m_f^2\right)\left( \left(m_P^2-2 m_P q_1^0 -m_S^2\right)^2-4m_f^2m_S^2 \right)}}{2 \left(m_f^2+m_P \left(m_P-2 q_1^0 \right)\right)} ,
\end{align}
and 
\begin{align}
 m_f < q_1^0 < \frac{m_P^2 -m_S^2 - 2m_f m_S }{2 m_P}.
\label{eq:q1_int_range}
\end{align}
By integrating the differential decay rate
\begin{align}
d\Gamma = \frac{1}{2p^0}\overline{|\mathcal{M}|^2}dQ 
 = \frac{1}{8(2\pi)^3 m_P}\left[\overline{|\mathcal{M}|^2}\delta\left(\cdots-\cos\theta_{13}\right)d\cos\theta_{13}\right]dq_1^0 dq_3^0 ,
\end{align}
for the range (\ref{eq:q3_int_range}) and (\ref{eq:q1_int_range}), we obtain the travel distance shown in Fig.~\ref{Fig:ctau}.

%



\end{document}